\documentclass[11pt,preprint]{aastex}

\shorttitle{Analytical analysis of Lyot coronographs}
\shortauthors{Ferrari}

\begin{document}

\title{Analytical analysis of Lyot coronographs response}

\author{Andr\'e Ferrari}
\affil{LUAN, Universit\'e de Nice Sophia-Antipolis,
    Parc Valrose, 06108 Nice cedex 02, France}

\begin{abstract}
We derive an analytical solution to the computation of the output of a 
Lyot coronagraph for a given complex amplitude on the pupil plane. This solution,
which does not require any simplifying assumption, relies on an expansion of the 
entrance complex amplitude on a Zernike base. According to this framework,
the main contribution of the paper is the expression of the response of the
coronagraph to a single base function. This result is illustrated by a computer
simulation which describes the classical effect of propagation of 
a tip-tilt error  in a coronagraph.

\end{abstract}

\keywords{instrumentation: adaptive optics -- techniques: high angular resolution 
}

\section{Introduction}

The discovery of extrasolar planets is at the origin of a renewed interest in
stellar coronagraphy. Considering the ambition of the targeted objectives, many authors
have pointed out the necessity for a very accurate analysis of the system in order
to study various undesired effects. For example, the specific properties of the light intensity measured by a system based on 
an extreme adaptive optics system and a coronagraph are the result that 
neither the residuals of the turbulence, nor the ideal coronagraphed point-spread function can be neglected with respect to the faint object (planet).
\citet{aimeithd04}  analyzed the fact that
the wavefront amplitudes associated  to these two contributions
will interfere leading to the so-called ``pinned'' speckles.
Another example is given by \citet{lloyd05} which pointed  out that a small
misalignment of the star with the center of the stop can result in a fake source. 
A related problem is also present in
\citep{soummer2005APJ} wich derives  the optimal apodization for an
arbitrary shaped aperture  using an algorithm proposed independently in
\citep{guyon2000} which relies on iterated simulations of the coronagraph response.

More generally, an intense activity aims to optimize the different 
coronagraph parameters (mask size, apodization shape,...) for a number of projects dedicated to devise  high-dynamic range imaging on the VLT (Sphere),  Gemini (GPI)
or the Subaru telescope (HiCIAO), see  for example \citep{IAUC200}.
The input/output relation of a coronagraph is in this case simulated by numerical
computations based on discrete Fourier transforms. 
However, such a numerical technique suffers from the well-known problems related 
to the choice of the extent of the sampled surface and the sampling frequency which both
define the sampling in the transformed domain. Note that this compromise is coupled
with the difficulty to evaluate numerically the  simulation errors.

This work focuses on the analytical characterization of the  response of a Lyot coronagraph. The objective is obviously also  to gain deeper insight in the behaviour of the system.
This problem has already been studied in the literature and analytical results were obtained under various assumptions.  In the one-dimensional case,  \citet{lloyd05} assume  that the Lyot stop  is band limited and the phase on the telescope aperture is small.
This last hypothesis is removed in \citep{sivara05} where the computation 
is carried for a rectangular pupil  assuming again that the Lyot stop is band-limited.
The development presented herein for a circular pupil differs from these approaches
substituting these simplifying assumptions by an expansion
of the  complex amplitude on an orthogonal basis.

 Section 1 recalls the general formalism of Lyot coronagraphy
 and justifies the choice of an expansion of the   complex amplitude on a Zernike base. 
 Section 2 contains the main results of the paper; the response of the 
 coronagraph to a Zernike polynomial is computed. The result involving an infinite sum, a bound on the truncation error is then derived. Section 3 presents two simulations.
First the response of the coronagraph to the 6 first Zernike functions is computed.
Then the formalism derived in this paper is used to illustrate the effect of a tip-tilt
error in a coronagraph.  A short appendix containing the
material required for the mathematical derivations of section 2 is included  at the end of the paper.

\section{Notations and hypothesis}
\label{secNotaandHyp}

\subsection{Coronagraph formalism}

We follow the notations of \citet{art_aa01b} and \citet{aim03}.
The successive planes  of the coronagraph 
are denoted by $A$, $B$, $C$ and $D$. $A$ is the entrance aperture, $B$ denotes the focal plane
 with the mask (without loss of generality we assume that the amplitude of the mask is $1-\epsilon$ where $\epsilon=1$ corresponds to the classical Lyot coronagraph and  
 $\epsilon=2$ to the   Roddier coronagraph), $C$ is the image  of the aperture with
 the Lyot stop and $D$  is the image in the focal plane after the coronagraph. 
The aperture transmission function is $p(x,y)$ and the wavefront complex amplitude in $A$
is $\Psi(x,y)$. In the case of an apodized pupil, we assume that the
apodization function is included in $\Psi(x,y)$.
In order to simplify the notations, the mask function  in $B$ is defined with coordinates proportional to  $1/\lambda f$ and decomposed as:
\begin{equation}
1 -\epsilon m\left(\frac{x}{\lambda f},\frac{y}{\lambda f}\right)
\end{equation}
where  function $m(.)$ equals to 1 inside the coronagraphic mask and 0 outside.

We will make in the sequel the usual approximations of paraxial optics.
Moreover we neglect the quadratic phase terms associated with the propagation 
of the waves or assume that the optical layout is properly designed to cancel it
\citep{aimeithd03}.
The expression in cartesian coordinates of the
complex amplitude in the successive planes are:
\begin{eqnarray}
\Psi_A(x,y)&=&\Psi(x,y)  p(x,y)  \\
\Psi_B(x,y)&=&\frac{1}{\jmath \lambda f} \widehat{\Psi_A}\left(\frac{ x}{\lambda f},\frac{ y}{\lambda f}\right)\left( 1 -\epsilon m\left(\frac{x}{\lambda f},\frac{y}{\lambda f}\right)\right) 
\label{planCcart}\\
\Psi_C(x,y)&=&\frac{1}{\jmath \lambda f} \widehat{\Psi_B}\left(\frac{ x}{\lambda f},\frac{ y}{\lambda f}\right)
p(-x,-y)   \\
&=&-  \left( \Psi_A(-x,-y) - \epsilon   
\left[\Psi_A(-u,-v)\ast \widehat{m}\left( u,v\right)\right]
\left(x,y\right) \right)p(-x,-y)  \label{geneplanC}\\
\Psi_D(x,y) &=&\frac{-1}{\jmath \lambda f}  
 \widehat{\Psi_A}\left(\frac{- x}{\lambda f},\frac{- y}{\lambda f}\right) \nonumber\\
 && \qquad
 +\epsilon  \frac{1}{\jmath \lambda f}  
 \left(
 \widehat{\Psi_A}(-x,-y)m(-x,-y)
 \ast 
 \widehat{p}(-x,-y)
 \right)
 \left(\frac{ x}{\lambda f},\frac{ y}{\lambda f}\right) \label{geneplanD}
 \end{eqnarray}
where $\hat{f}$ is the Fourier transform of $f$ and $\ast$ denotes convolution.
Eqs.~(\ref{geneplanC},\ref{geneplanD}) assume that the Lyot stop is the same
as the pupil. However  for classical ``unapodized'' 
Lyot coronagraph the residual intensity in plane $C$ is concentrated
 at the edges of the pupil and a reduction of the Lyot stop size
is needed in order to improve the rejection. 
The case of a reduced Lyot stop, which consists in convolving
Eq.~(\ref{geneplanD}) by the appropriate function, has not been considered
in Eqs.~(\ref{geneplanC},\ref{geneplanD}) to alleviate the notations but  will be discussed in section 3.
 It is important to note that the reduction of the Lyot stop can be avoided
using a prolate apodized entrance pupil which will optimally concentrate
the residual amplitude in $C$, see for example \citep{art_aa01b}.

The coronagraph response being derived herein for a circular pupil, the use of polar coordinates
will be preferred. 
Transcription of previous equations to polar coordinates is straightforward.
Moreover, as long as the aperture transmission function and the stop have a circular symmetry, their Fourier transform will verify the same symmetry, as proved by
Eq.~(\ref{hankelgeneral}) with $m=0$, i.e. the Hankel transform.
This leads  to the following expression of the complex amplitude in~$D$:
\begin{equation}
\Psi_D(r\lambda f,\theta)=\frac{-1}{\jmath \lambda f}  
\widehat{\Psi_A}( r, \theta+\pi) 
+\frac{\epsilon }{j \lambda f}
\left(
\left(\widehat{\Psi_A}( r,\theta+\pi)m(r)\right)
\ast
\widehat{p}(r)
\right)
(r,\theta) \label{ampcompD}
\end{equation}
where the convolution of the two functions is still computed with respect to 
to the cartesian coordinates $(x,y)$.

\subsection{Choice of a base}

As mentioned in the introduction,  the analytical 
computation of the coronagraph response proposed herein relies on the 
expansion of the  complex amplitude in $A$ on an orthogonal basis.
Eq.~(\ref{ampcompD}) shows that the coronagraph \emph{acts linearly} on the complex amplitude, consequently the problems simplifies to the computation of the response of 
each basis function.
The retained solution consists in the expansion of the  complex amplitude in $A$
on Zernike polynomials. Basic  properties of the Zernike polynomials required in the paper
 are recalled in appendix \ref{appendA}.

Adopting the usual ordering of the Zernike circle polynomial
\citep{maha94} we can write:
\begin{eqnarray}
\Psi_A(r,\theta) &=&  \sum_{(m,n)} a_{(m,n)} U_n^m(r/R,\theta)\label{expanA0}\\
&=& \sum_k a_k Z_k(r/R,\theta),\;a_k\equiv a_{(m,n)}\in \mathbb{C} \label{expanA}
\end{eqnarray}
where $R$ is the radius of the aperture.
This expansion is rather unusual, the Zernike polynomials
being generally used for the expansion of the wavefront. However
it is worthy to note that, as Eq.~(\ref{geneplanC}) shows, 
a coronagraphic system will always introduce amplitude aberration. 
Hence, even in the case of a perfect wave with no aberration in $A$, 
an expansion of only the phase in $C$ will not be appropriate.
Finally, Eq.~(\ref{expanA}) can also be justified by the fact that it coincides (up to a linear
transform) with the classical approximation  of the complex amplitude in the case of  sufficiently small  phase errors assuming a first order development of the exponential function.

We will illustrate the expansion (\ref{expanA}) in the case of tip-tilt error
with an apodized pupil:
\begin{equation}
\Psi_A(rR,\theta) = a(r)\Pi(r)e^{\jmath \beta r \cos(\theta)  }
\end{equation}
where $a(r)$ denotes the pupil apodization and $\Pi(r) = 1$ for $r\in [0,1)$ 
and $0$ if $r\geq 1$.
Computation of the  projection of $\Psi_A(r,\theta)$  on $U_n^m(r/R,\theta)$ is
straightforward using the definition of the Bessel functions  of integer order 
\citep{abramb}:
\begin{eqnarray}
\int_0^{2\pi} \int_0^R \Psi_A(r,\theta) U_n^m(r/R,\theta) r dr d\theta &=&R^2\int_0^1\int_0^{2\pi} R_n^m(r) \cos(m\theta) a(r) e^{\jmath \beta r \cos(\theta)  } r dr d\theta \\
&=&2\pi R^2 \jmath^m \int_0^1 a(r)R_n^m(r)J_m(\beta r)rdr
 \label{apotilt}
\end{eqnarray}
The projection of $\Psi_A(r,\theta)$  on $U_n^{-m}(r/R,\theta)$ equals 0.

\begin{itemize}
\item In the unapodized case, $a(r)=1$, integral in Eq.~(\ref{apotilt}) can be computed
using Eq.~(\ref{intbesselradial}):
\begin{equation}
2\pi R^2 \jmath^m \int_0^1 a(r)R_n^m(r)J_m(\beta r)rdr=2\pi R^2 \jmath^m (-1)^\frac{n-m}{2}\frac{J_{n+1}(\beta)}{\beta}
\end{equation}
The coefficient $a_{k}$ is then obtained dividing this quantity by the $L^2$ norm of the 
Zernike polynomials \citep{bornb}, leading to:
\begin{equation}
a_k = \jmath^m  (-1)^\frac{n-m}{2} \frac{4(n+1)}{1+\delta(m)}\frac{J_{n+1}(\beta)}{\beta}
\label{amntilt}
\end{equation}

\item A particularly important case is that where $a(r)$ is proportional to 
the circular prolate function $\varphi_{0,0}(c,rR)$, \citep{aim03}. In this
case the  integral in  Eq.~(\ref{apotilt}) can be computed using the
expansion of  $\varphi_{0,0}(c,r)$ derived in  \citep{slep64}:
\begin{equation}
\varphi_{0,0}(c,r) = \sum_{k=0}^\infty d_k^{0,0}(c) \sqrt{r} F(k+1,-k;1;r^2)
\end{equation}
The function $F(k+1,-k;1;r^2)$ defined in Eq.~(\ref{hyperg}) reduces
to a polynomial of order $2k$ which, as mentioned in  \citep{slep64} ``is 
closely related to the Zernike polynomials''. 
Indeed using Eq.~(\ref{zernger}) and the results below it
can be easily checked that: $F(k+1,-k;1;r^2) = (-1)^k R_{2k}^0(r)$. Inserting this expansion in Eq.~(\ref{apotilt})
and integrating terms by terms leads  to integrals which
generalize  Eq.~(\ref{intbesselradial}). These integrals can be computed 
for example using of integrals of the  type 
$\int_0^1 r^{\nu} J_m(\beta r)dr$ \citep{gradb}. This derivation
will not be presented herein for sake of brevity.

\end{itemize}

Finally, for more complicated complex amplitudes, the  $a_k$ 
can be of course computed numerically. This problem as been addressed in
\citep{pawl02} using a piecewise approximation of $\Psi_A(x,y)$ over
a lattice of squares with size $\Delta \times \Delta$ and 
centered on point $(x_i,y_j)$. In this case the estimation of $a_{k} $ is given by:
\begin{equation}
\hat{a}_{k} = \sum_{(x_i,y_j) \in \mathcal{D}} \Psi_A(x_i,y_j)w_n^m( x_i,y_j)^\ast
\label{eqpawl}
\end{equation}
where $w_n^m(x_i,y_j)$ is the integral of the Zernike polynomial $U_n^m(\rho/R,\phi)$
over the square centered on $(x_i,y_j)$. \citep{pawl02} gives bound for the
mean integrated squared error on the reconstruction of $\Psi_A(x,y)$ when
the coefficients are given by Eq.~(\ref{eqpawl}). This analysis is particularly important in
our case because it quantifies the dependence of the error on the smoothness 
of $\Psi_A(x,y)$, the sampling rate $\Delta$ and the  geometrical error due to the 
circular geometry of the pupil.

\section{Coronagraph response}

\subsection{Response of the coronagraph to a Zernike polynomial}

The purpose of this section is to compute the complex amplitude
in $D$ when the complex amplitude in $A$ is the Zernike polynomial with
radial degree $n$ and azimuthal frequency $m$. In this case
the complex amplitude $\Psi_D(r,\theta)$ will be denoted as
$\mathcal{D}_n^m(r,\theta)$.
According to Eq.~(\ref{ampcompD}), the difficulty in the computation of
$\mathcal{D}_n^m(r,\theta)$ lies  in the evaluation  of the convolution:
\begin{equation}
\Xi(r,\theta)=
\left(
\left(\widehat{\Psi_A}( r,\theta+\pi)m(r)\right)
\ast 
\widehat{p}(r)
\right)
(r,\theta) \label{convol1}
\end{equation}
In this expression  $m(r)$ is an ``annular'' mask of radius 
$d$ which, with the definition adopted in Eq.~(\ref{planCcart}) is   defined as:
\begin{equation}
m(r) = \Pi\left( r \frac{\lambda f}{d} \right) \label{theMask}
\end{equation}

\begin{figure}
\begin{center}
\includegraphics[width=.6\textwidth]{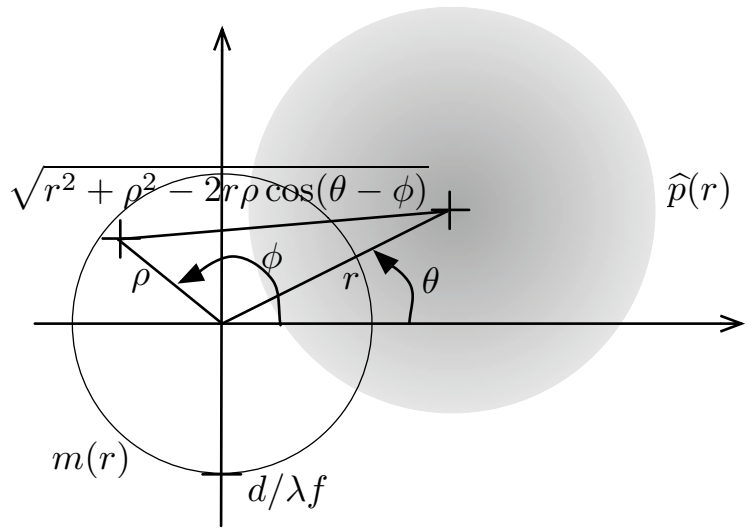}
\end{center}
\caption{Computation of the convolution between $\widehat{\Psi_A}( r,\theta+\pi)m(r)$ and
$\widehat{p}(r)$. \label{compconv}}
\end{figure}

The computation of the convolution in $\Xi(r,\theta)$ is sketched in Fig.~\ref{compconv}. Using Eq.~(\ref{theMask}),  $\Xi(r,\theta)$ simplifies to:
\begin{equation}
\Xi(r,\theta)= \int_0^{d/\lambda f }\int_0^{2\pi}
\widehat{\Psi_A}(\rho,\phi+\pi)
\widehat{p}\left(\sqrt{r^2+\rho^2-2r\rho \cos(\theta-\phi)} \right) \rho d\rho d\phi 
\end{equation}
The next step consists in substituting in this equation:
\begin{itemize}
\item $\widehat{p}(r)$ by the Fourier transform of $p(r) = \Pi\left(r/R\right)$:
\begin{equation}
 \hat{p}(r)=\frac{RJ_1(2\pi R r)}{r} \label{Airy}
\end{equation}
\item $\Psi_A(\rho,\phi)$ by $U_n^m(\rho/R,\phi)$ and
consequently  $\widehat{\Psi_A}(\rho,\phi)$ by $R^2\widehat{U_n^m}(rR,\phi)$ where
$\widehat{U_n^m}(r,\phi)$ is given in Eq.~(\ref{fourier1zern}).
\end{itemize}

In order to simplify the notations we define the new ``standardized'' integral $\tilde{\Xi}(r,\theta,\xi)$ by:
\begin{equation}
\tilde{\Xi}(r,\theta,\xi)=
\int_0^\xi \int_0^{2\pi}
\cos(m\phi)J_{n+1}(\rho)
\frac{J_1\left(
\sqrt{r^2+\rho^2-2r\rho \cos(\theta-\phi)} \right)}
{\sqrt{r^2+\rho^2-2r\rho \cos(\theta-\phi)}} 
d\rho d\phi \label{convol2}
\end{equation}
It can be easily checked in this case that:
\begin{equation}
\Xi(r,\theta)= R^2 \jmath^{m} (-1)^\frac{n-m}{2} 
\tilde{\Xi}\left(2\pi R r,\theta,\frac{2\pi R d}{\lambda f} \right) \label{xixitild}
\end{equation}

Analytical computation of $\tilde{\Xi}(r,\theta,\xi)$ relies on the properties of the Gegenbauer
polynomials defined in appendix \ref{appendA}. Substituting Eq.~(\ref{eqGegen}) 
for $\nu=1$ in  Eq.~(\ref{convol2}) allows indeed to separate the integrations with respect to
$\rho$ and $\phi$:
\begin{equation}
\tilde{\Xi}(r,\theta,\xi)=
2\sum_{k=0}^\infty (k+1) \frac{J_{k+1}(r)}{r}
\int_0^\xi \frac{J_{k+1}(\rho)J_{n+1}(\rho)}{\rho}d\rho
\int_0^{2\pi}
\cos(m\phi) C_k^{(1)}(\cos(\theta-\phi)) d\phi \label{convol3}
\end{equation}

\begin{itemize}
\item Computation of the integral on $\phi$ is straightforward using (\ref{cheby}):
\begin{displaymath}
\int_0^{2\pi} \cos(m\phi) C_k^{(1)}(\cos(\theta-\phi))d\phi = \pi \cos(m\theta)
 \sum_{q=0}^k \delta(m-k+2q) 
\end{displaymath} 
\item  Computation of the integral on $\rho$ relies on recursion formulas on
indefinite integrals of products of Bessel functions, \citep{abramb}:
 \begin{eqnarray}
&& k\not= n,\;\int_0^\xi \frac{J_{n}(\rho)J_{k}(\rho)}{\rho} d\rho=
   \frac{\xi J_{k-1}(\xi)J_{n}(\xi)-\xi J_{k}(\xi)J_{n-1}(\xi) ) +(n-k)
 J_{n}(\xi)J_{k}(\xi) }{k^2-n^2} \label{int1} \\
&& \int_0^\xi \frac{J_{n}(\rho)^2}{\rho} d\rho=
 \frac{1}{2n}(1-J_0(\xi)^2 -2\sum_{q=1}^{n-1}J_q(\xi)^2
 -J_n(\xi)^2)  \label{int2} 
 \end{eqnarray} 
 
 \end{itemize}

After computation of the integral  of Eq.~(\ref{convol3}), substitution of 
Eq.~(\ref{xixitild}) in Eq.~(\ref{ampcompD}) gives the complex amplitude in $D$ for a single basis function
$\Psi_A( r,\theta)= U_n^m(r/R,\theta)$:
\begin{equation}
  \mathcal{D}_n^m( r,\theta)= 
\jmath^{m-1}  (-1)^\frac{n-m}{2}R \cos(m\theta) 
 \left( -\frac{J_{n+1}(2\pi \mu r)}{r}  
+ \epsilon  \sum_{k=0}^\infty 
\eta_{m,n,k}(2\pi \mu d)
\frac{J_{k+1}(2\pi \mu r )}{r} \right) \label{leresultat}
\end{equation}
with $\mu = R/{\lambda f}$ and:
\begin{equation}
\eta_{m,n,k}(\xi) =
(k+1)
 \big(\sum_{q=0}^k \delta(m-k+2q) \big)\int_0^\xi \frac{J_{n+1}(\rho)J_{k+1}(\rho)}{\rho} d\rho \label{defeta}
\end{equation}

The corresponding complex amplitude in $C$ for $r<R$ can be
directly computed from  Eq.~(\ref{leresultat}) using the inverse Fourier transform
of $\cos(m\theta)J_{k+1}(2\pi r)/r$ obtained in Eq.~(\ref{zernger}):
\begin{equation}
  \mathcal{C}_n^m(r,\theta)= 
  (-1)^\frac{n-m}{2} \cos(m\theta) 
 \left( -\mathcal{R}_n^m\left( \frac{r}{R}\right) 
+ \epsilon  \sum_{k=0}^\infty 
\eta_{m,n,k}(2\pi \mu d)
\mathcal{R}_k^m\left( \frac{r}{R}\right) \right)  \label{leresultat2}
\end{equation}

Eqs.~(\ref{leresultat},\ref{leresultat2}) give an analytical expression of the complex amplitude in $C$ for $r<R$ and in $D$ when a single basis function is applied in $A$
and when the size of the Lyot stop equals the size of the entrance pupil.
In the general where the amplitude in $A$ is given by Eqs.~(\ref{expanA0},\ref{expanA}),
the complex amplitudes in $C$ and $D$ become:
\begin{equation}
\Psi_C(r,\theta) = \sum_{(m,n)} a_{(m,n)} \mathcal{C}_n^m(r,\theta),\;
\Psi_D(r,\theta) = \sum_{(m,n)} a_{(m,n)}  \mathcal{D}_n^m(r,\theta) \label{repcomplete}
\end{equation}

As mentioned in section 2.1, if the entrance pupil is not apodized
a reduction of  the Lyot stop must be considered.  This is achieved
replacing   $p(r)$ by  $p(\alpha^{-1}r)$ with  $\alpha<1$.
  The expression of the complex amplitude in 
$C$ is of course  straightforward and for example Eq.~(\ref{leresultat2}) becomes
$\mathcal{C}_n^m(r,\theta) p(\alpha^{-1}r)$. This result allows numerical
computation of the complex amplitude in $D$ using a 
single Fourier transform.
Unfortunately  it is much more complicated to obtain 
an analytical expression of the complex amplitude in $D$. 
The derivation presented above can be of course
redeveloped  replacing $\hat{p}(r)$ by $\alpha^2 \hat{p}(\alpha r)$ and
straightforward computation shows that:
\begin{enumerate}
\item Similarly to  Eq.~(\ref{leresultat}), the convolution (\ref{convol1})
will expand in an infinite sum of  functions $\cos(m\theta)J_{k+1}(2\pi \alpha \mu r )/r$. 
However, the ``radial contribution'' to the coefficients weighting these functions, 
see Eq.~(\ref{defeta}),  becomes:
\begin{displaymath}
\int_0^\xi \frac{J_{k+1}(\rho)J_{n+1}(\alpha^{-1}\rho)}{\rho}d\rho
\end{displaymath}
which  cannot be computed straightforwardly
as in Eqs.~(\ref{int1},\ref{int2}).

\item The first term in Eq.~(\ref{ampcompD}) is now replaced by the Fourier
transform of $U_n^m(\rho/R,\phi) \Pi(r/\alpha R)$ which cannot be anymore
calculated using Eq.~(\ref{intbesselradial}).
\end{enumerate}

\subsection{Bound for the truncation error of $\mathcal{D}_n^m(r,\theta)$}

As we are interested in the computation of 
 $\mathcal{C}_n^m(r,\theta)$ or
$\mathcal{D}_n^m(r,\theta)$ 
from the implementation of formula (\ref{leresultat}),
the errors produced when the infinite sum is truncated must be studied.
In order to reduce mathematical developments
we only present herein the results for   $\mathcal{D}_n^m(r,\theta)$
when the size of the Lyot stop equals the size of the pupil.

We define the truncation error on $\mathcal{D}_n^m(r,\theta)$:
\begin{equation}
\mathcal{E}_N(r,\theta;m,n,\mu,d) = \epsilon R  \left|\cos(m\theta) 
\sum_{k=N+1}^\infty 
\eta_{m,n,k}(2\pi \mu d)
\frac{J_{k+1}(2\pi \mu r)}{r}
\right|
\end{equation}

Computation of a bound on the truncation error relies on the 
classical upper bound for the  Bessel functions of integer order \citep{abramb}:
\begin{equation}
\left| J_{k+1}(r)\right| \leq  \frac{(r/2)^{k+1}}{k!},\; r\geq 0 \label{majorbess}
\end{equation}
Substitution of this result in Eq.~(\ref{defeta}) gives:
\begin{eqnarray}
\eta_{m,n,k}(\xi)  &\leq& (k+1)
 \big(\sum_{q=0}^k \delta(m-k+2q) \big) \frac{1}{k+n+2}\frac{1}{k!n!}
 \left(\frac{\xi}{2} \right)^{k+n+2}\\
 & \leq &  \frac{k+1}{k!n!}
 \left(\frac{\xi}{2} \right)^{k+n+2}
\end{eqnarray}
which leads to the following bound for the truncation error:
\begin{equation}
\mathcal{E}_N(r,\theta;m,n,\mu,d) \leq
\frac{\epsilon R(\pi \mu)^{3+n}d^{2+n}}{n!}
\sum_{k=N+1}^\infty \frac{k+1}{(k!)^2}\left( (\pi \mu)^2 rd \right)^k \label{laborne}
\end{equation}

The above serie  is absolutely convergent for $r>0$. 
As a consequence the expansion in Eq.~(\ref{leresultat}) converges uniformly 
for $(r,\theta)\in [0,\infty)\times [0,2\pi)$. Finally, it is worthy to note that
the computation of the infinite sum in the upper bound (\ref{laborne}) can be avoided 
using the equality:
\begin{equation}
\sum_{k=0}^\infty \frac{k+1}{(k!)^2}x^k =
I_0(2\sqrt{x})+\sqrt{x}I_1(2\sqrt{x}) 
\end{equation}
where $I_\nu(x)$ is the modified Bessel function.

\section{Simulation results}

\subsection{Response of the coronagraph to the first Zernike function}

\begin{figure}
\includegraphics[width=.9\textwidth]{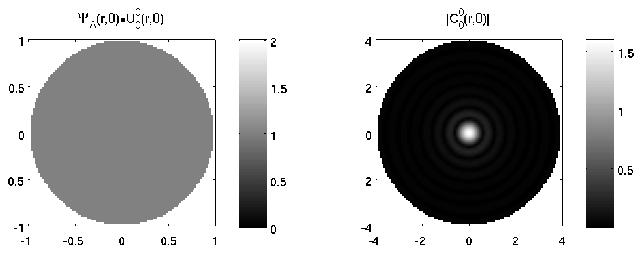}\\
\includegraphics[width=.9\textwidth]{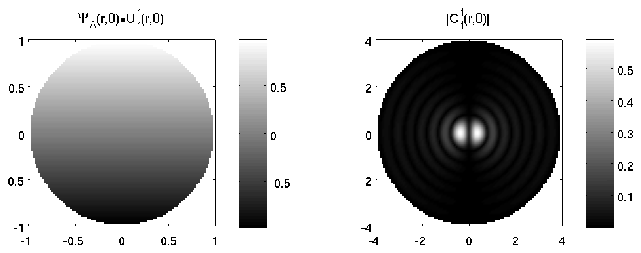}\\
\includegraphics[width=.9\textwidth]{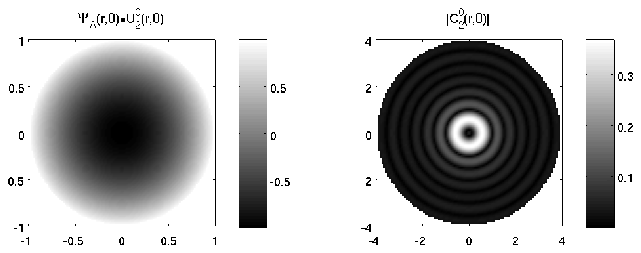}\\
\caption{Complex amplitude in $A$ and squared root of the amplitude in $D$, i.e.
$|\mathcal{D}_n^m(r,\theta)|$. 
The parameters used in the simulation are: $\lambda f =1$, $R=1$, $d=3$, $\epsilon = 1$
(Lyot coronagraph). \label{figresu1}}
\end{figure}

\begin{figure}
\includegraphics[width=.9\textwidth]{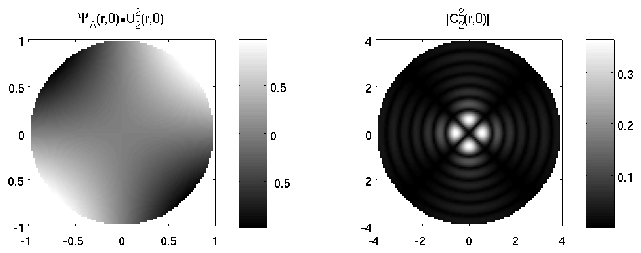}\\
\includegraphics[width=.9\textwidth]{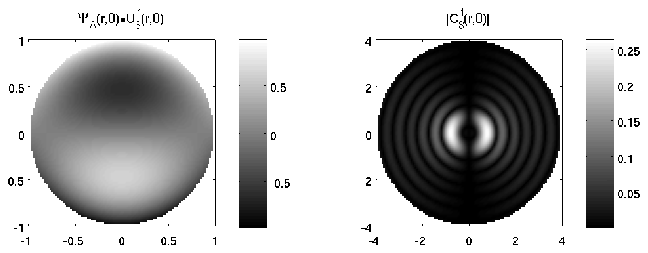}\\
\includegraphics[width=.9\textwidth]{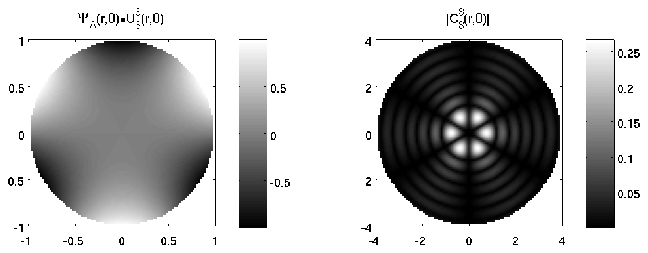}\\
\caption{Complex amplitude in $A$ and squared root of the amplitude in $D$, i.e.
$|\mathcal{D}_n^m(r,\theta)|$. 
The parameters used in the simulation are: $\lambda f =1$, $R=1$, $d=3$, $\epsilon = 1$
(Lyot coronagraph).  \label{figresu2}}
\end{figure}

Figures \ref{figresu1} and \ref{figresu2} give the intensity in the $D$ plane
of the coronagraph when the complex amplitude in the $A$ plane is one of
the first six Zernike polynomials. The complex amplitudes have been computed
using Eq.~(\ref{leresultat}). Each raw contains  
$U_n^m(r,\theta)$ and $\mathcal{D}_n^m(r,\theta)$
for a given couple $(n,m)$. These plots have been obtained truncating
the infinite summation of Eq.  (\ref{leresultat}) to the first 40 terms.

\begin{figure}
\begin{center}
\includegraphics[width=.8\textwidth]{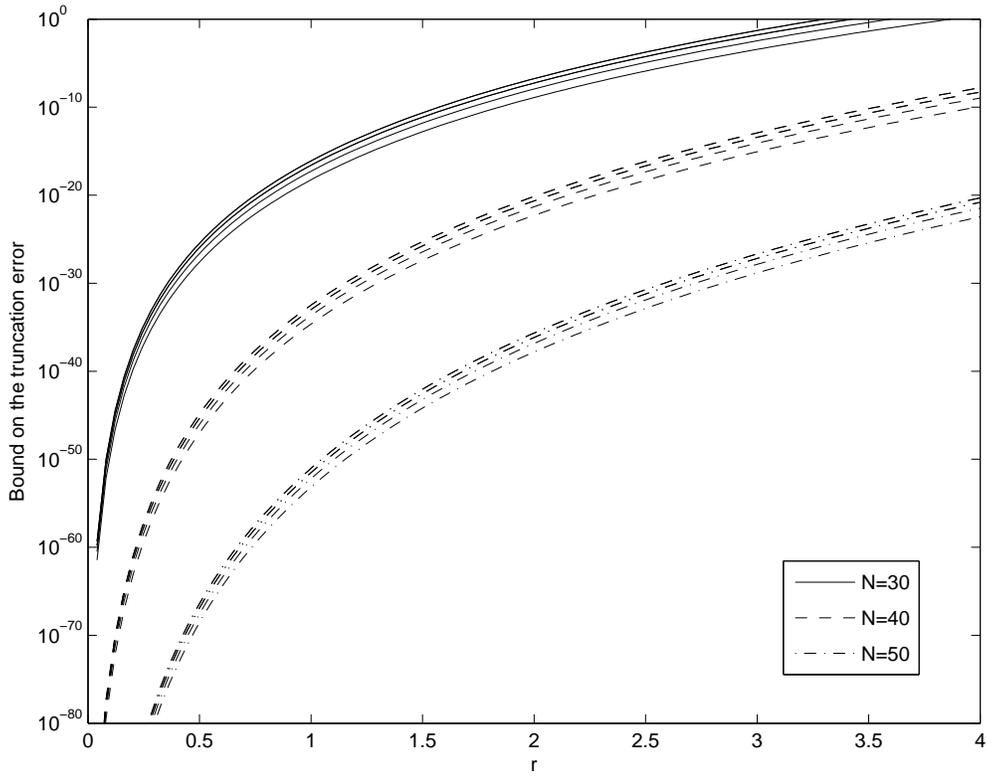}
\end{center}
\caption{Bounds on the truncation error as a function of $r$. The parameters
are the same as the parameters used for Figs. \ref{figresu1} and \ref{figresu2}.
For each value of $N$, the bound is plot for the first 6 Zernike polynomials. 
\label{plotbound}}
\end{figure}

The relevance of the truncation error bound is verified in 
Fig.~\ref{plotbound}. This plot shows the error bound (\ref{laborne})
as a function of $r$ for the parameters used in Figs.  \ref{figresu1} and \ref{figresu2}.
The increase of the bound with $r$ is simply due to the fact that the majoration
of $|J_{k+1}(r)|$ given by Eq.~(\ref{majorbess}) is only relevant for small values of $r$
as long as $|J_{k+1}(r)|$ is bounded on $[0,\infty)$.
It is important to note that this plot justifies,  at least for this configuration, 
the validity of a  truncation to  $N=40$ for the computation of  
$\mathcal{D}_n^m(r,\theta)$. In this case the truncation error is in fact always
less than $10^{-10}$.

\subsection{Application to tip-tilt error analysis}

The effects of a tip-tilt error in Lyot coronagraphs has been extensively studied
by  \citet{lloyd05} and \citet{sivara05}. The scope of the simulation presented here
is only to validate
the results derived in section 2 simulating the particular case 
where there is a misalignment of the  star with the center of the stop.
According to the previous notations the complex amplitude in $D$
decomposes as Eq.~(\ref{repcomplete}). In the case of a  tip-tilt error in $A$, 
the values of the coefficients $a_{(m,n)}$ are given  by Eq.~(\ref{amntilt}).

\begin{figure}
\begin{center}
\includegraphics[width=.45\textwidth]{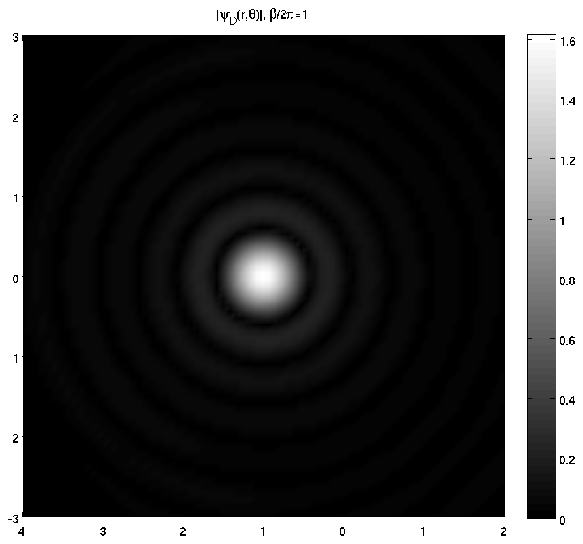}\quad
\includegraphics[width=.45\textwidth]{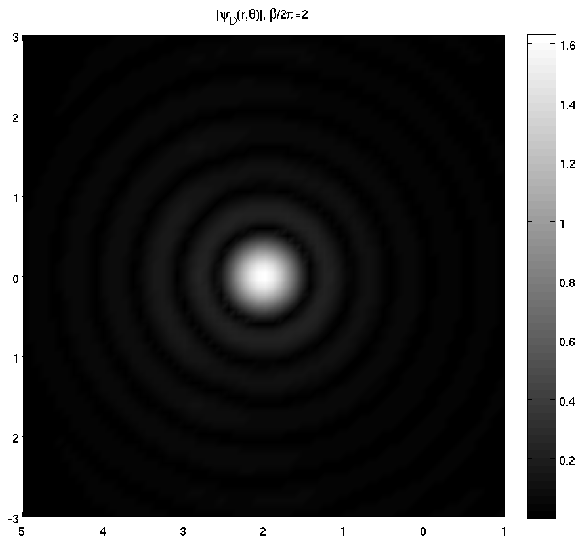}\\
\includegraphics[width=.45\textwidth]{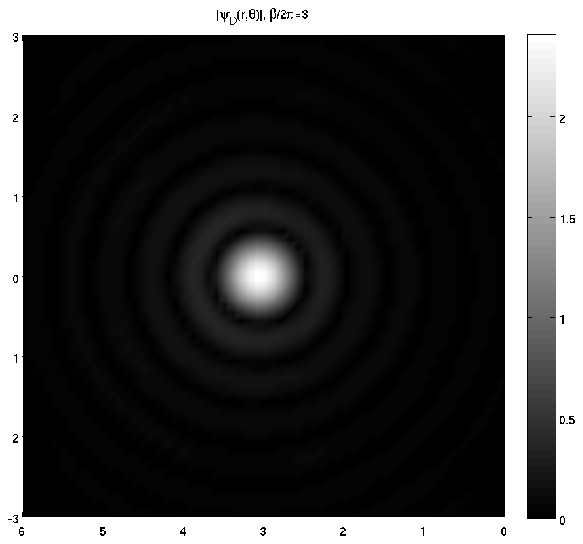}\quad
\includegraphics[width=.45\textwidth]{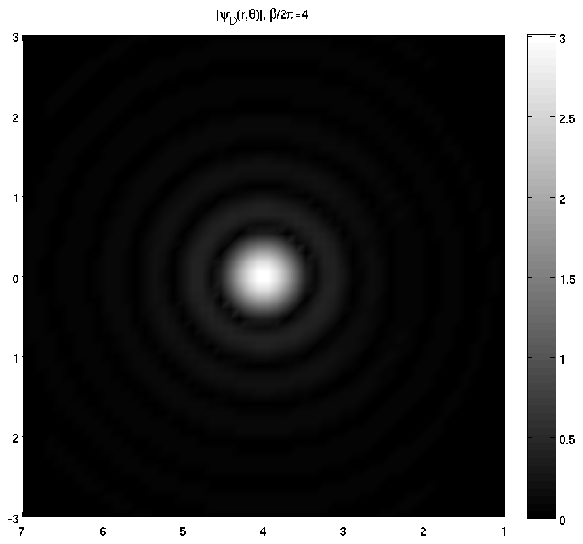}
\end{center}
\caption{$|\Psi_D(r,\theta)|$ for different values of $\beta$. The parameters
are the same as the parameters used for Figs. \ref{figresu1} and \ref{figresu2}.
\label{plottilt}}
\end{figure}

Fig.~\ref{plottilt} shows $|\Psi_D(r,\theta)|$ for different values of $\beta>0$
(the case $\beta=0$ is given in the first row of Fig.~\ref{figresu1}).
The truncation in the summation (\ref{repcomplete}) has been chosen
taking into account that Eq.~(\ref{amntilt}) implies:
\begin{displaymath}
|a_{(m,n)}| \sim  \frac{ 4}{\sqrt{2\pi}\beta (1+\delta(m))}
\sqrt{n}
\left( \frac{e\beta}{2n}\right)^n,\; \mbox{when}\; n\rightarrow \infty
\end{displaymath}
Note that according to the notations of Eq.~(\ref{expanA}), $\Psi_B(r,\theta)$
equals Eq.~(\ref{Airy}) shifted of $-\beta\lambda f/(2\pi R)$ on axis $x$.
Consequently,  the star is behind the focal stop in the first two images and
outside in the last one.

\section{Conclusion}

In this paper we have presented a theoretical formalism for the analytical study of
the Lyot coronagraph response.  The main purposes of this work
are of course to assist coronagraph design  but also to
improve data processing performances
for the detection and characterization of  extrasolar planets.
\begin{itemize}
\item The first application  is the computation of the
response of the coronagraph to a planet at a given position. 
This is achieved for example in the case of
a classical Lyot coronagraph using Eqs.~(\ref{repcomplete},\ref{amntilt}).
This point is essential for the derivation of an optimal decision scheme to test
the presence of a planet at a given location.

\item This formalism can also be applied to fully characterize  the statistical properties of the
complex amplitude in the $D$ plane. For a given spatial covariance in $A$ which is
fixed through the covariance of coefficients $a_k$, 
the spatial covariance in $D$ becomes:
\begin{equation}
{\mathsf{cov}}[\Psi_D(r,\theta) \Psi_D(r',\theta') ] = 
 \sum_{k,l} \mathsf{cov}[a_k,a_l]  \mathcal{D}_k(r,\theta)
 \mathcal{D}_l(r',\theta')
\end{equation}

Although detection algorithms  based solely on the marginal distribution
of the complex amplitude can be developed as in \citep{IAUC05a},
the use of an accurate model for the spatial correlation of the complex amplitude is 
essential  in order to derive detection algorithms with optimal performances,
 as demonstrated in \citep{icassp06}.
\end{itemize}

\acknowledgments 

The author thanks the anonymous  referee who helped improve the paper.
The author is also grateful to Claude Aime and R\'emi Soummer for helpful discussions and insightful comments.

\appendix
\section{Appendix}
\label{appendA}

This section presents some facts about  Fourier transform in polar coordinates, Zernike and Gegenbauer polynomials.

Among the various available possibilities to define an orthogonal set of functions on the unit radius disk a central position is hold by the Zernike polynomials, see for example
\citep{maha94} and included references. They are defined  for $n\geq m $ by:
\begin{equation}
U_n^m(r,\theta) = R_n^m(r) \cos(m\theta)\Pi(r),\;
U_n^{-m}(r,\theta) = R_n^m(r) \sin(m\theta)\Pi(r) \label{leszern}
\end{equation}
when  $n$ et $m$   share the same parity.
The $R_n^m(r)$ are the radial polynomials. Different normalizations exist for 
$R_n^m(r)$, we retain herein the definition of \citep{bornb}: $R_n^m(1)=1$. 
Among many properties verified by these polynomials, we focus on:
\begin{equation}
\int_{0}^{1} r R_n^m(r)  J_m(v r) dr = 
(-1)^\frac{n-m}{2}\frac{J_{n+1}(v)}{v} \label{intbesselradial}
\end{equation}
see \citep[appendix VII]{bornb} for the proof.
This equality allows straightforward computation of the Fourier transform of the Zernike polynomials. In fact recall first that when $f(r,\theta)=g(r)\cos(m\theta)$, $m\in \mathbb{Z}$, a simple
change of variables in the Fourier transform integral leads to:
\begin{equation}
\hat{f}(\rho,\phi)= 2\pi (-\jmath) ^m  \cos(m\phi) 
\int_{0}^{\infty} r g(r)  J_m(2\pi r \rho) dr \label{hankelgeneral}
\end{equation}
An analog result for the inverse Fourier transform of 
$\hat{f}(\rho,\phi)=h(\rho)\cos(m\phi)$ is:
\begin{equation}
f(r,\theta)= 2\pi \jmath ^m  \cos(m\theta) 
\int_{0}^{\infty} \rho h(\rho)  J_m(2\pi r \rho) d\rho \label{hankelinvgeneral}
\end{equation}

Applying the result of Eq.~(\ref{hankelgeneral}) with Eq.~(\ref{intbesselradial}) immediately gives:
\begin{eqnarray}
&&\widehat{U_n^m}(\rho,\phi) =  \jmath^m (-1)^\frac{n+m}{2} \cos(m\phi)\frac{J_{n+1}(2\pi \rho)}{\rho} \label{fourier1zern}\\
&&\widehat{U_n^{-m}}(\rho,\phi) =  \jmath^m (-1)^\frac{n+m}{2} \sin(m\phi)
\frac{J_{n+1}(2\pi \rho)}{\rho}
\end{eqnarray}

The previous equation gives the inverse Fourier transform of
$\cos(m\phi)\frac{J_{n+1}(2\pi \rho)}{\rho}$  
when $n\geq m\geq 0$ and
 $n$ et $m$   share the same parity.
In the general case where $n\geq 0$ and $m\geq 0$
this inverse Fourier transform, denoted as $f(r,\theta)$ 
must be computed independently. 
If we subsitute $h(r)$ by $J_{n+1}(2\pi \rho)/\rho$ in 
Eq. (\ref{hankelinvgeneral}) the resulting integral is a Weber-Schafheitlin
type integral \citep{abramb}. This results in
$f(r,\theta) =\jmath^m\cos(m\theta) \mathcal{R}_n^m(r)$ where: 
\begin{equation}
 \mbox{if}\;r<1, \; \mathcal{R}_n^m(r)=
r^m
 \frac{\Gamma\left( \frac{n+m}{2} +1 \right)}
 {\Gamma(m+1)\Gamma\left( \frac{n-m}{2} +1 \right)}
F\left( \frac{n+m}{2} +1,\frac{m-n}{2} ;m+1,r^2\right) \label{zernger}
\end{equation}
$F(a,b;c;z)$ is the Gauss hypergeometric function,  see  \citep{gradb}:
\begin{equation}
F(a,b;c;z) = 1 + \frac{ab}{1!c}z+ \frac{a(a+1)b(b+1)}{2!c(c+1)}z^2+\cdots
 \label{hyperg}
\end{equation}
It is interesting to note from Eqs.~(\ref{zernger}) and (\ref{hyperg}) that
if $b=(m-n)/2 \in \mathbb{Z}^-$ the sum in Eq.~($\ref{hyperg}$) reduces to a
polynom in $z$ of order $-(m-n)/2$. Consequently $\mathcal{R}_n^m(r)$ reduces
to a polynom with degree $n$ which of course coincides up to 
$(-1) ^{(m-n)/2}$ with $R_n^m(r)$ for $r\leq 1$. For this reason
$\mathcal{R}_n^m(r)$ can be considered as a natural generalization of the
Zernike polynomials. Note that, contrarily to the generalization proposed in
\citep{myri66} or \citep{wuns05}, this generalization is not a polynomial.

We now briefly give the principal results related to the Gegenbauer polynomials.
See for example  \citep{SpecialFunctions} or \citep{abramb} for detailed properties.
The Gegenbauer (or ultraspherical) polynomials, noted as $t \mapsto C_k^{(\nu)}(t)$
are defined as the coefficients  of the power series expansion of 
$r \mapsto (1-2rt+r^2)^{-\nu}$:
\begin{displaymath}
\frac{1}{(1-2rt+r^2)^\nu} = \sum_{k=0}^\infty C_k^{(\nu)}(t)r^k
\end{displaymath}
For example $ C_k^{(1)}(t)$ gives the Chebyshev polynomial of the second kind  $U_k(t)$:
\begin{equation}
C_k^{(1)}(\cos(\psi))=\sum_{q=0}^k \cos((k-2q)\psi) \label{cheby}
\end{equation}

Among the numerous  beautiful properties of the Gegenbauer polynomials, 
we focus on the expansion:
\begin{equation}
\frac{J_\nu(w)}{w}=
  2^\nu  \Gamma(\nu) \sum_{k=0}^\infty
(k+\nu )  \frac{J_{k+\nu}(r)}{r^\nu} \frac{J_{k+\nu}(\rho)}{\rho^\nu} C_k^{(\nu)} (\cos(\gamma))
\label{eqGegen}
\end{equation}
where $w=\sqrt{r^2+\rho^2-2r\rho \cos(\gamma)}$.

\end{document}